# The Er doping of ZnCr$_2$O$_4$

Chen Yang[1,2], Danrui Ni[1], Nan Yao[3], and Robert J. Cava[1*]

[1] Department of Chemistry, Princeton University, Princeton, New Jersey, 08544

[2] Department of Electrical and Computer Engineering, Princeton University, Princeton, New Jersey, 08544

[3] Princeton Materials Institute, Princeton University, Princeton, NJ 08544, USA

*corresponding author's email: cy11@princeton.edu

**Abstract**

Magnetic Er$^{3+}$ is doped into the well-studied frustrated normal spinel ZnCr$_2$O$_4$. Various spectroscopies are employed to prove that Er$^{3+}$ successfully enters the spinel to form ZnCr$_{2-x}$Er$_x$O$_4$ for x less than 0.005. The low levels of Er$^{3+}$ doping possible nonetheless have a significant effect on the frustrated magnetism and the ordering that is seen near 12 K in the undoped material.

**Introduction**

ZnCr$_2$O$_4$ is a well-studied AB$_2$O$_4$ normal spinel and is one of the prototype materials for the study of frustrated Heisenberg magnetism[1]. It has a face-centered cubic crystal structure within which the Cr$^{3+}$ B-sites form a Kagome lattice on 111 layers (**Figure 1 inset**) and a magnetic lattice of corner-sharing tetrahedra when the equally distant next-plane Cr sites are considered. In previous studies on solid solutions of ZnCr$_2$O$_4$ and spinels such as MgCr$_2$O$_4$ and CdCr$_2$O$_4$, the magnetic lattice has been modified by doping with transition metals such as Mn, Fe, and Co to study novel catalytic and magnetic behavior[2–5]. Distortion on the A site is observed if the dopant successfully substitutes for the original nonmagnetic A site ions[6,7]. By substituting a magnetic ion for a nonmagnetic ion, the magnetic pinning effect sets a preferred spin orientation in these types of spinel solid solutions[4,7]. However, current studies are still lacking the modification of spinel solid solutions obtained by doping a magnetic ion on the B site. Since rare-earth ions with unpaired 4*f* electrons have very different magnetic behavior from 3*d* transition metal ions, our goal in the current study is to see how rare-earth substitution for Cr alters the spin glass magnetism in a normal spinel.

$Er^{3+}$ (S=15/2), a small rare-earth but a relatively large cation, can potentially substitute for $Cr^{3+}$ (S=3/2) on the Kagome net ([8,9]). The fact that $Er^{3+}$ has a strong preference for octahedral over tetrahedral coordination in oxides has been described in many previous works[10–13]. Thus, in the present work we get heterogenous Kagome nets mostly composed of $Cr^{3+}$ with a very small amount of $Er^{3+}$ included as a magnetic disturbance. We have successfully synthesized polycrystalline samples with nominal chemical formulas: $ZnCr_{2-x}Er_xO_4$ with $x$=0, 0.002, and 0.004. The successful doping of Er in $ZnCr_2O_4$ at this low level is demonstrated via different diffraction techniques and spectroscopies. We study the changes in structural and magnetic frustration in the doped samples. Structural analysis was done by X-ray diffraction. Vibrational analysis was studied by Raman and IR spectroscopies. We hope to inspire studies on frustrated magnetism in a partially disturbed Kagome lattice with two types of magnetic ions.

**Experimental**

*Polycrystalline Synthesis*: The $Er_2O_3$ employed (Thermo scientific 99.99%) was dried at 900 ºC in air overnight. The starting materials ZnO, $Cr_2O_3$ and $Er_2O_3$ were weighed in the appropriate molar ratios. The resulting powder was mixed well and transferred into alumina crucibles. The samples were heated in a high-temperature furnace (Sentro Tech Corp. ST-1600C-445 High-Temperature Box Furnace) in air at 1300 ºC and 1500 ºC for 24 hours at each temperature. Powder X-ray diffraction (PXRD) measurements were carried out after each grinding step utilizing a Brucker D8 FOCUS diffractometer with Cu Kα radiation ($\lambda_{K\alpha}$ = 1.5406 Å).

*Magnetic Property Measurements:* Magnetic measurements on single-phase polycrystalline samples were performed using a Quantum Design Dynacool Physical Property Measurement System (PPMS). The magnetic susceptibility ($\chi$) was defined as the ratio of the magnetization (M) to the applied field (H). Zero-field cooled (ZFC) magnetic data were acquired under a magnetic field of H = 0.1 T (1000 Oe) over a temperature range spanning from 1.8 K to 300 K. A modified Curie-Weiss law $\chi - \chi_0 = C/(T - \theta_{CW})$ was fitted to the inverse $\chi$ data over the temperature range of 225-275 K in order to derive the Curie constant (C) and Weiss temperature ($\theta_{CW}$), with $\chi_0$ representing a temperature-independent correction factor accounting for the core diamagnetism and the magnetization of the sample holders. The effective magnetic moment ($\mu_{eff}$) per formula unit, expressed in Bohr magnetons (μ$_B$), was calculated from the constant C in the above expression using √8C. Isothermal magnetization under magnetic fields ranging from -9 T to +9 T (-90.000 to + 90,000 Oe) at 1.8K and 300K was collected.

Changes in magnetic susceptibility were observed. Thus, in addition to the X-ray and magnetism measurements, Raman spectroscopy, IR, SEM, EDX, and $^3$He heat capacity (HC) characterization experiments were used to verify that the Er doping was successful.

**Results and Discussion**

The parent phase of ZnCr$_2$O$_4$ was reported to be face-centered cubic with space group *Fd-3m* (227) with Zn in tetrahedral sites and Cr in octahedral sites[14]. The intersecting Cr layers form a Kagome lattice in such a crystal structure, the origin of the geometric magnetic frustration[8], shown in Figure 1 inset. The PXRD patterns of ZnCr$_{2-x}$Er$_x$O$_4$ (**Figure 1**) did not indicate any clear peak shift for increasing *x* at the low doping levels employed. However, a low-intensity impurity peak belonging to Er$_2$O$_3$ was observed for *x*=0.005. This leads to the conclusion that the highest doping concentration possible is less than *x*=0.005.

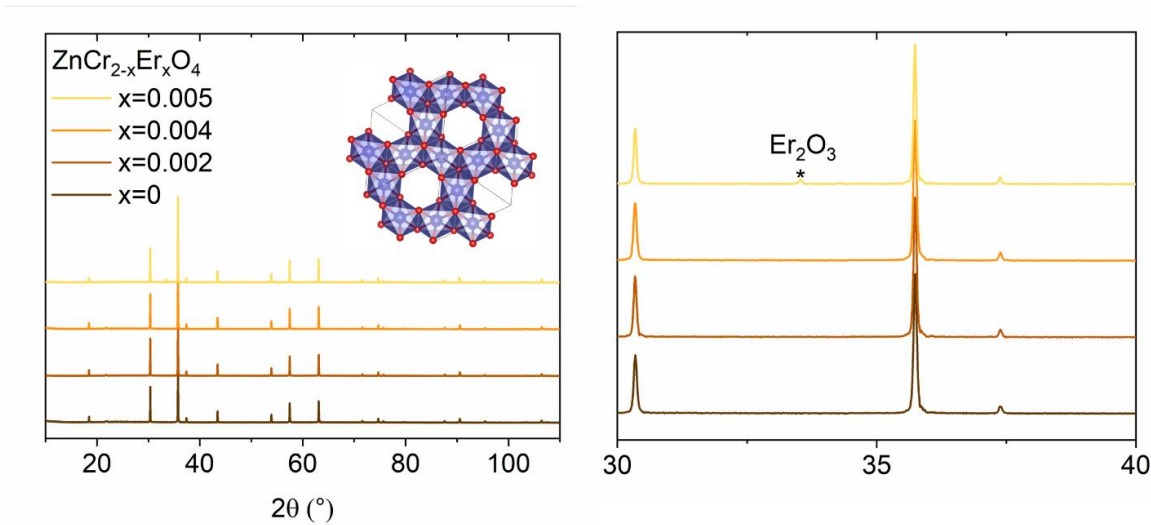

**Figure 1. PXRD patterns for ZnCr$_{2-x}$Er$_x$O$_4$ with *x*=0, 0.002, 0.004 and 0.005. The inset on the left is a CrO6 Kagome net in a normal spinel structure. The right figure shows an expanded region of the PXRD pattern.**

Raman microscope spectroscopy, as another common way to investigate structural change, is more sensitive to small doping concentrations. A Raman microscope (Thermofisher DXR3xi) collected spectra at room temperature (**Figure 2**). Samples were pressed into a pellet and sintered at 1500 ºC for 8 hours. This assured significant signal intensities and a good signal-to-noise ratio. All spectra were collected using a 50 μm pinhole and 10x objective. The excitation laser was unpolarized 532nm, favoring Er ([15]). **Figure 2** shows the Raman spectra taken on pellet samples of $x$=0, 0.002, and 0.004, and a reference scan for a dry $Er_2O_3$ pellet sample under the same laser power, exposure time, and number of scans from the 0~2000 $cm^{-1}$ range. The maximum peak of $Er_2O_3$ is absent across all doped spinel samples, proving that there is no $Er_2O_3$ impurity present. $ZnCr_2O_4$ has Raman-active phonon modes at 185, 514, 607, and 691 $cm^{-1}$ with the labelled peaks matched to the reported results in green[16]. Tetrahedral and octahedral site-based Raman peaks have been reported around 400$cm^{-1}$ and 900$cm^{-1}$ [17]. In the Figure, emerging peaks are clearly observed for x = 0.004 around 400$cm^{-1}$ and 900$cm^{-1}$ (highlighted in blue in **Figure 2**). The change in Raman patterns can be taken as convincing evidence of successful Er doping. Er-O could induce symmetry breaking, octahedral distortion and change in the crystal field[18]. A vacancy peak has also been reported around 941$cm^{-1}$, but it is absent in our $x$=0 and 0.002 samples; it is, however, observed for our $x$=0.004 sample. Further study of the Raman spectrum would be of interest.

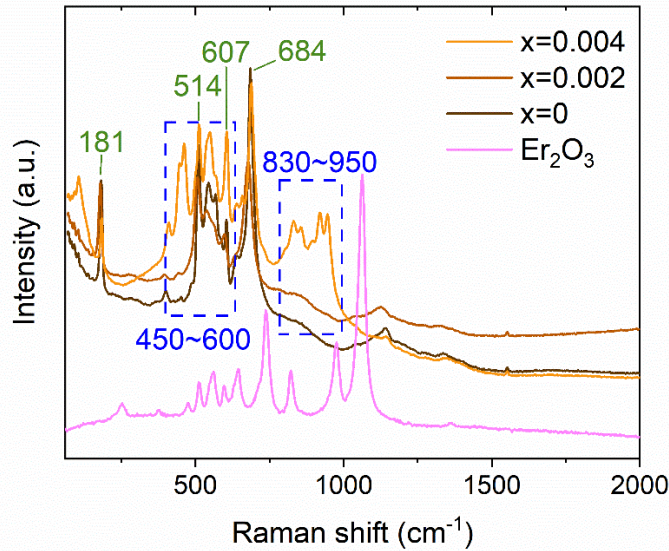

**Figure 2. Raman spectra. The excitation wavelength was 532nm. The green labeled peaks indicate the peaks that match the reported phonon modes. The blue dashed areas are highlighted to indicate the emergent peaks.**

SEM images and EDX spectra were taken on a Verios 460 XHR SEM to closely check the powder uniformity and the presence or absence of trace amounts of $Er_2O_3$ (**Figure 3**). Under the same magnification, the SEM showed similar particle shapes and uniformity on pellet samples of $x=0$ (left) and $x=0.004$ (right). (The horizontal streaks were caused by charging of the insulating samples.) EDX spectra were collected using a 15 keV beam energy (**Figure S1**). The atomic percentages were collected to trace the amount of Er in $x=0$ and $x=0.004$ samples, showing that the percent of Er present was approximately 0 and 0.5, respectively. This is further evidence to prove the successful doping of Er into the spinel.

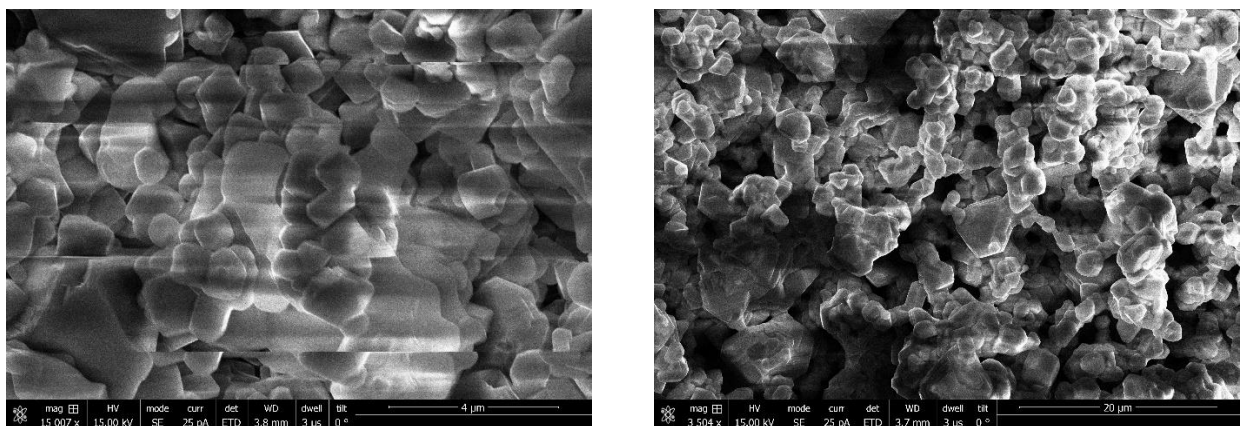

**Figure 3. SEM images taken under 15 keV beam energy. Left is the undoped sample and right is the *x*=0.004 sample. Both samples were in pellet form sintered under the same temperature**.

Attenuated-Total-Reflection (ATR)-IR spectra were also collected on very small amounts of loose powder samples. Transmittance vs. wavenumber ranging from 400 cm$^{-1}$ to 4000 cm$^{-1}$ was collected by subtracting the background (**Figure 4**). Two principal absorption peaks, for Cr-O and Zn-O, located at 490cm$^{-1}$ and 615cm$^{-1}$ [19,20] were observed. The introduction of $Er^{3+}$ reduced the absorption at the principal absorption bands. In previous IR studies, 1100cm$^{-1}$ is a common C-O mode[21]. The absence of a 1630cm$^{-1}$ peak indicated nonexistence of O-H bonds[19–21] in our samples.

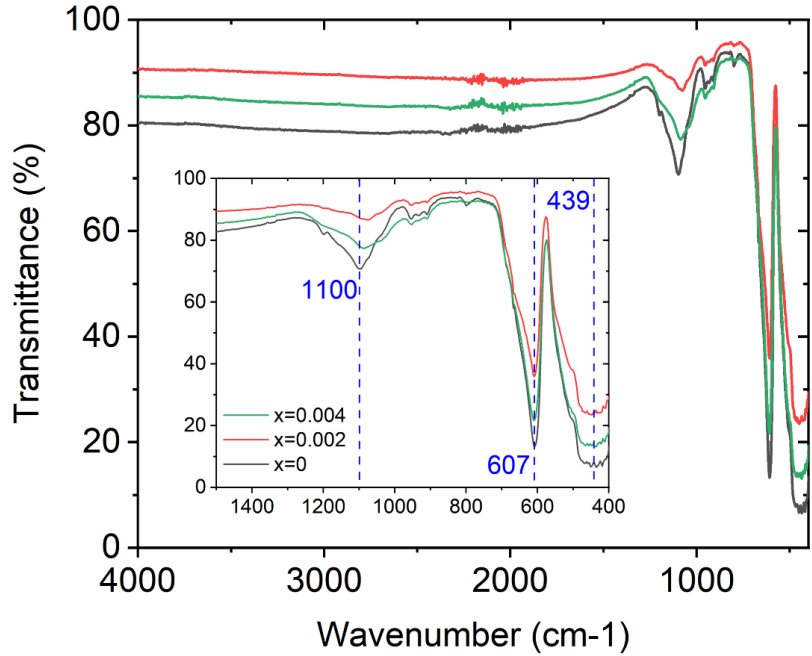

**Figure 4. The ATR-IR spectra taken on all the single phase samples. The inset is an expansion that shows most peak features.**

DC magnetic susceptibility data was obtained on samples of $x=0$, 0.002, and 0.004 (**Figure 5(a) and (c)**). To help to determine whether it was the size or the magnetic moment of Er that induced the changes in the $1/\chi$ curves, we synthesized nonmagnetic Y and Al doped versions of $ZnCr_{2-x}M_xO_4$ for $x=0.004$. $Er^{3+}$, $Y^{3+}$, and $Al^{3+}$ in octahedra have ionic radii of 0.89Å, 0.9Å, and 0.39Å, respectively. In the zoom-in figure, it is shown that the magnetic frustration hump did not change its transition temperature or shape. Therefore, we can draw the conclusion that the magnetic frustration was suppressed by the $Er^{3+}$ magnetic moment instead of the dopant size. Curie-Weiss fitted parameters are shown in **Table 1**. $ZnCr_2O_4$ was reported to be a Heisenberg-type antiferromagnetic which has $T_N$ = 12.5K, $\theta_{cw}$ = -390K, $\mu_{eff}$ = 3.89 $\mu_B$ for antiferromagnetic $ZnCr_2O_4$. The large and negative value of $\theta_{cw}$ is an indication of strong negative super-exchange of S=3/2 Cr spins. ([22–24]). Our $T_N$ = 12.3K and $\mu_{eff}$ = 3.875(8) $\mu_B$ are close to the reported value. However, our $\theta_{cw}$ is smaller than reported and can be caused by several possible reasons. Firstly, oxygen deficiency brought down $T_N$ since our reaction temperature was higher than people's reported experimental recipes ([8,25]). Secondly, Zn and Cr partial site mixing was reported previously, where, it was shown that $T_N$ decreased as $ZnCr_2O_4$ became more nonstoichiometric ($^{26,27}$). Given that the magnetic moment of $Er^{3+}$ = 9.58 µB ($^9$) and $Cr^{3+}$ = 3.87 µ$_B$ ($^8$), the effective moment should increase with the introduction of $Er^{3+}$ and the experimental results confirm with trend in **Table 1**. This supports the successful doping of Er. Isothermal magnetization results are shown in **Figure 7**. The M vs. H curves for $ZnCr_2O_4$ are poorly reported. Neither sample was magnetically saturated up to an applied field of 9T. In the Er-doped sample, the competition between $Er^{3+}$ and $Cr^{3+}$ moments change the curve shape.

In the zoom-in temperature in **Figure 5 (b) and (d)**, it is seen that the phase transition features got weaker and became subtle in the $x$=0.004 sample. Yet we could still see a weak transition at around 10K. Considering that we synthesized the doped sample at a much higher than the reported temperature of 1100 °C ($^{26}$), we would like to study further on how temperature will affect the magnetic behavior (**Figure 6**). The lower temperature stoichiometric sample was synthesized at 1100 °C. The result shows that the transition temperature happened at a slightly higher temperature compared to the 1500 °C sample. This suggested higher temperature might cause vaporization of Cr ions and thus resulted in defects in the samples.

**Table 1. Curie-Weiss fitting parameters for Figure 5.**

|  | Fit Range (K) | Fitted $\theta_{cw}$ (K) | Experimental $\mu_{eff}$ ($\mu_B$/f.u.) |
|---|---|---|---|
| $ZnCr_2O_4$ | 225-275 | - 346.22(1) | 5.385(1) |
| $ZnCr_{1.998}Er_{0.002}O_4$ | 225-275 | - 353.97(2) | 5.953(1) |
| $ZnCr_{1.996}Er_{0.004}O_4$ | 225-275 | - 435.50(9) | 5.858(1) |
| $ZnCr_{1.996}Al_{0.004}O_4$ | 225-275 | - 366.75(6) | 5.174(9) |
| $ZnCr_{1.996}Y_{0.004}O_4$ | 225-275 | - 311.64(2) | 5.130(3) |

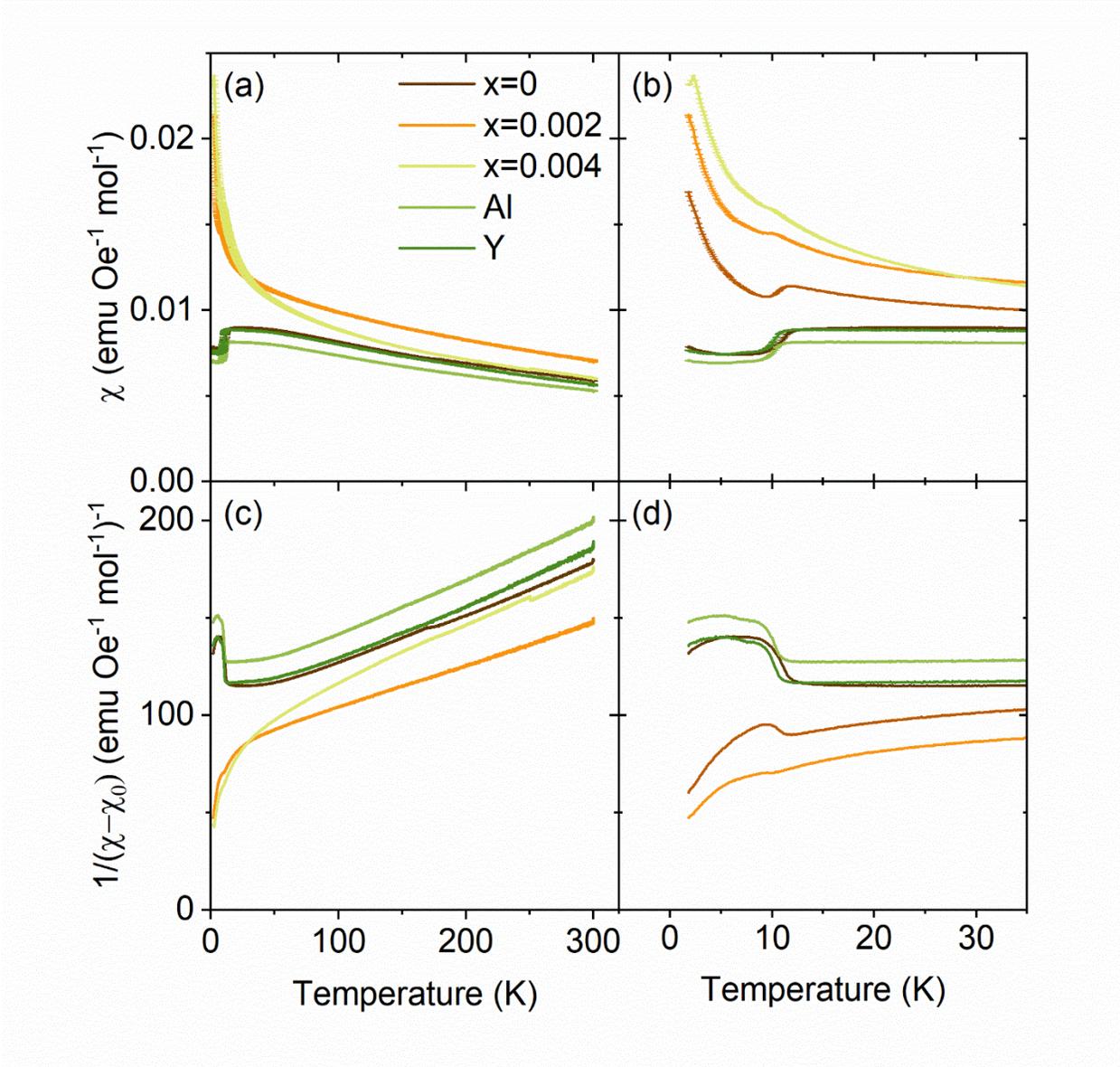

**Figure 5. The temperature dependent DC magnetic susceptibility for different doping and preparation conditions.**

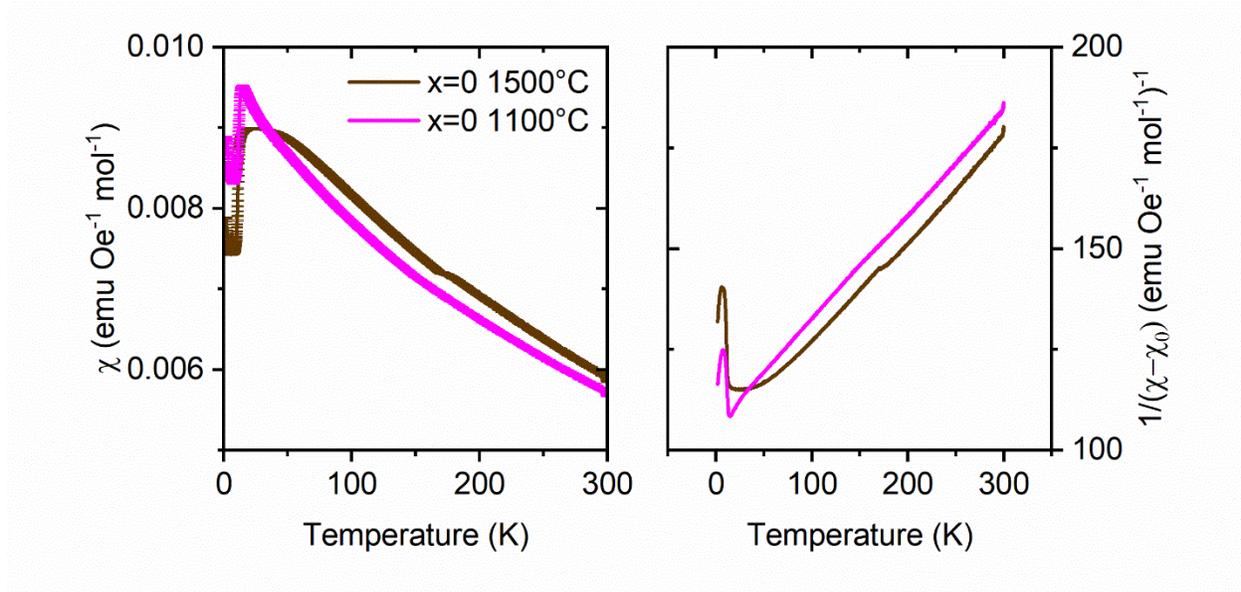

**Figure 6.** The temperature dependent DC magnetic susceptibility of ZnCrO4 prepared in air at different temperatures.

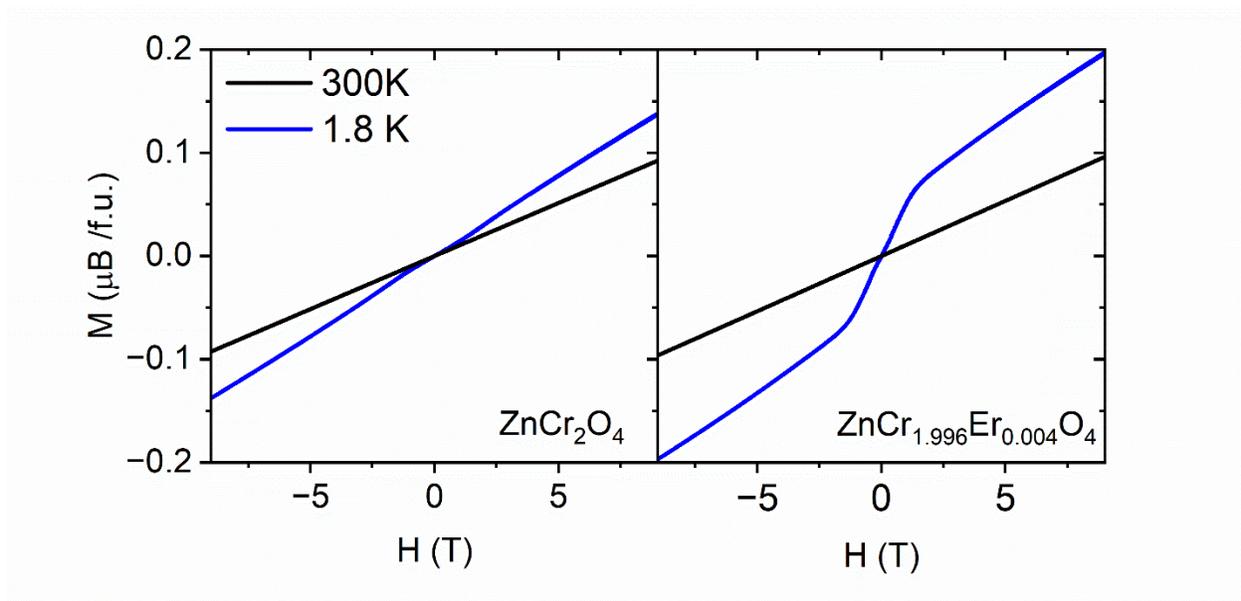

**Figure 7.** The isothermal magnetization for the x=0 and x=0.004 samples at 1.8K and 300K.

To further characterize the materials, specific heat capacity data were collected from 0.5~20K under 0 and 1 T (**Figure 8**). The doped sample has a lower transition temperature than the undoped one, indicating that the $Er^{3+}$ moments delay the long-range ordering to lower temperature. In **Figure 8**, we used an equal-area method to estimate transition temperatures to be around 12.5K and 10.5K for undoped and doped samples respectively, which are close to the transition temperatures seen in our magnetic susceptibility data., The blue-dashed lines are best considered as guides for the eye. We also compared the FWHM of the peaks under 0T of applied field, the peak of the doped sample is half as wide as hat of the undoped sample. The application of 1T magnetic field did not change the FWHM of the curves. For the undoped sample, we suspected that the actual transition peak is between T=10~11.10K since the peak tip was not clear.

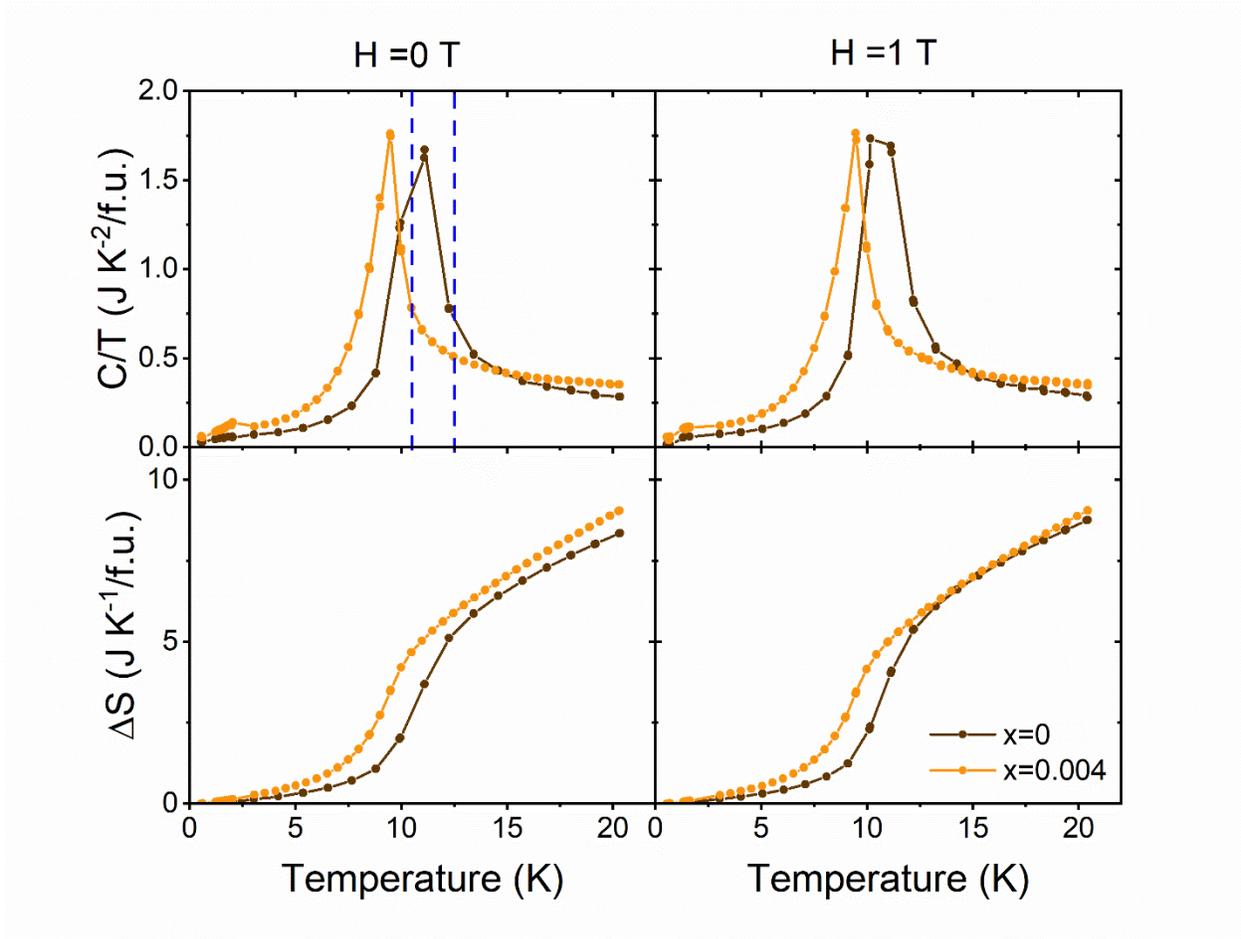

**Figure 8. Specific heat capacity data for *x*=0 and *x*=0.004 under 0T and 1T magnetic fields. (Blue-dashed lines are guides to the eye)**

**Conclusions**

We have shown that it is possible to dope small amounts (less than 0.5%) of a highly magnetic ion into the normal spinel $ZnCr_2O_4$. Even the small amount of doping found results in significant changes in the magnetic behavior of the material. Studies of larger and smaller dopants with no magnetic moment show that it is in fact the magnetism of the Er that impacts the behavior most, not the ionic size or disorder. Further, although it has been reported that higher temperature synthesis result in disorder of this material, this is not seen for the current case, Various follow up studies are suggested including one where many three-plus rare-earth ions are substituted for Cr in $ZnCr_2O_4$ to determine which has the greatest effect on the magnetism.

**Acknowledgments**

This material is based upon work supported by the U.S. Department of Energy, Office of Science, National Quantum Information Science Research Centers, Co-design Center for Quantum Advantage (C2QA) under contract number DE-SC0012704. The authors acknowledge the use of Princeton's Imaging and Analysis Center, which is partially supported by the Princeton Center for Complex Materials, a National Science Foundation (NSF)-MRSEC program (DMR-2011750). The research at Michigan State University was funded by the U.S. Department of Energy, Basic Energy Sciences under contract number DE-SC0023568.